\documentclass[12pt]{article}
\usepackage{graphicx}
\usepackage{epsfig}
\usepackage{epstopdf}
\DeclareGraphicsExtensions{.pdf,.eps,.png,.jpg,.mps}

\def\lsim{\mathrel{\rlap{\lower3pt\hbox{\hskip0pt$\sim$}}
     \raise1pt\hbox{$<$}}}         
\def\gsim{\mathrel{\rlap{\lower4pt\hbox{\hskip1pt$\sim$}}
     \raise1pt\hbox{$>$}}}         

\usepackage{amsmath}
\usepackage{amsfonts}

\begin{document}
\begin{titlepage}

\centerline{\Large \bf On Origins of Bubbles}
\medskip

\centerline{Zura Kakushadze$^\S$$^\dag$\footnote{\, Zura Kakushadze, Ph.D., is the President of Quantigic$^\circledR$ Solutions LLC,
and a Full Professor at Free University of Tbilisi. Email: \tt zura@quantigic.com}}
\bigskip

\centerline{\em $^\S$ Quantigic$^\circledR$ Solutions LLC}
\centerline{\em 1127 High Ridge Road \#135, Stamford, CT 06905\,\,\footnote{\, DISCLAIMER: This address is used by the corresponding author for no
purpose other than to indicate his professional affiliation as is customary in
publications. In particular, the contents of this paper
are not intended as an investment, legal, tax or any other such advice,
and in no way represent views of Quantigic$^\circledR$ Solutions LLC,
the website \underline{www.quantigic.com} or any of their other affiliates.
}}
\medskip
\centerline{\em $^\dag$ Business School \& School of Physics, Free University of Tbilisi}
\centerline{\em 240, David Agmashenebeli Alley, Tbilisi, 0159, Georgia}
\medskip
\centerline{(August 27, 2016)}

\bigskip
\medskip

\begin{abstract}
{}We discuss -- in what is intended to be a pedagogical fashion -- a criterion, which is a lower bound on a certain ratio, for when a stock (or a similar instrument) is not a good investment in the long term, which can happen even if the expected return is positive. The root cause is that prices are positive and have skewed, long-tailed distributions, which coupled with volatility results in a long-run asymmetry. This relates to bubbles in stock prices, which we discuss using a simple binomial tree model, without resorting to the stochastic calculus machinery. We illustrate empirical properties of the aforesaid ratio. Log of market cap and sectors appear to be relevant explanatory variables for this ratio, while price-to-book ratio (or its log) is not. We also discuss a short-term effect of volatility, to wit, the analog of Heisenberg's uncertainty principle in finance and a simple derivation thereof using a binary tree.
\end{abstract}
\medskip
\end{titlepage}

\newpage

\section{Introduction}

{}Risk and reward are two integral parts of investing.\footnote{\, See, e.g., \cite{Sharpe1966}.} Thus, we have expected return (which we can reasonably expect at some future time by investing into a given instrument) based on some considerations rooted in data available to us now. If this expected return is positive and guaranteed, then it is a no-brainer: we invest and reap the reward. However, few things in life are guaranteed (barring death and taxes). Our investment in all likelihood bears some risk, which in some cases can be (gu)estimated based on historical data. The question is, is there a relationship between risk and return that can tell us whether an investment is a ``good" one?\footnote{\, Or, alternatively, that it is a ``bad" one.}

{}The Sharpe ratio \cite{Sharpe1994} provides a simple measure of such a relationship. Up to a normalization constant (see below) it is simply the ratio of the expected return over the volatility (or the historical standard deviation of realized returns). However, the Sharpe ratio depends on the time horizon for which it is calculated. E.g., daily expected return and volatility give us daily Sharpe ratio, which is on average lower (by a factor of $\sqrt{d}$, where $d\approx 252$ is the approximate number of trading days in a year, if we focus on stocks) than the annualized Sharpe ratio. If the daily expected return and volatility do not change much in time, then the Sharpe ratio goes to infinity as $\sqrt{T}$ with the time horizon $T$. So, a practical way of thinking about the Sharpe ratio is that, if, say, the annualized Sharpe ratio is 2, then the probability of losing money in a given year is less than about 2.3\% (assuming normally distributed realized returns, that is, which can be farfetched -- see below).\footnote{\, Recall the 68-95-99.7 rule: if $x$ is a normally distributed variable with mean $\mu$ and standard deviation $\sigma$, then we have the following probabilities: $\mbox{Pr}(\mu - n\sigma \leq x \leq \mu + n\sigma) = P_n$, $P_1\approx 68.27\%$, $P_2 \approx 95.45\%$, $P_3\approx 99.73\%$. The probability of losing money when the Sharpe ratio equals $n$ (i.e., $\mu = n\sigma$) then is ${\widetilde P} = (1 - P_n)/2$. So, we have ${\widetilde P}_1 \approx 15.9\%$, ${\widetilde P}_2 \approx 2.3\%$, ${\widetilde P}_3 \approx 0.14\%$. However, this does not take into account leverage, margin calls, investor withdrawals and other such nuances.}

{}So, is there a simple criterion which tells us, based on the expected return and volatility, if an investment is (not) worthwhile in the long term? The answer is yes, at least under some assumptions. It relates to what constitutes a bubble in the price of an instrument. Hereafter we focus on stocks for the sake of terminological concreteness. However, our discussion applies to a wider array of instruments.

{}There is a wealth of literature on bubbles in stock prices (see, e.g., \cite{Protter} and references therein), which utilizes a complex notion of strict local martingales and the stochastic calculus machinery. Here we deliberately avoid such complexities. Our aim is to discuss the issue as simply and intuitively as possible, and our discussion is intended to be pedagogical, so it is by no means a vigorous mathematical derivation of anything. We use a simple binary tree to illustrate our main point.

{}So, what is a stock price bubble? This is tricky. It depends, inter alia, on the time horizon. Let us focus on large time horizons. Then we have a bubble if at time $t=0$ the stock price $S_0$ is greater than its fair market price $S_*$ as $t\rightarrow\infty$. The question is, what is this fair market price $S_*$? This is where the volatility comes in.

{}Imagine, just for a second, that the stock price follows a random walk (or a Brownian motion).\footnote{\, This is not realistic because the price must be positive. But bear with us.} Then, if the expected value of the stock price is constant (i.e., the stock process is a martingale), holding the stock appears to be a rational thing to do: there is a 50-50 chance of making/losing money. However, stock prices are positive, which makes all the difference. These prices are not normally distributed, but have skewed, long-tailed distributions at the higher end. Thus, if the stock process follows a geometric Brownian motion\footnote{\, Alternatively, we can consider a discrete time version thereof.}, the prices are log-normally distributed. If the expected price is constant (a martingale), this no longer means that there is a 50-50 chance of making/losing money. In fact, the most {\em probable} long-run price in this case is {\em zero}! That is, investing into a stock with a constant expected value of its price is a sure way to lose money! The stock must have nonzero drift $\nu$ (whereby its expected value grows as $\exp(\nu t)$ at large $t$) for investing into the stock to make sense. The lower bound $\nu > \nu_*$ on the drift is given by $\nu_* = \sigma^2/2$. So, we can define the ratio
\begin{equation}
 \kappa = {2\nu\over\sigma^2}
\end{equation}
and we should {\em not} invest into (non-dividend-paying) stocks in the long term unless $\kappa > 1$. The ratio $\kappa$, which apparently has not been named in finance, is dimensionless and, unlike the Sharpe ratio, is independent of the time horizon (if $\nu$ and $\sigma$ are).

{}So, positiveness of stock prices, which (as is the case with many positive-valued quantities) results in a skewed distribution therefor, coupled with stock volatility creates a long-term asymmetry whereby ``on average" is not the same as ``most probably", and it is the latter that matters, albeit at times it may be the former that is followed (behaviorally). We discuss this using a simple binary tree model in Section \ref{sec2}, including for dividend-paying stocks.\footnote{\, And the na\"ive intuition that dividends do not make a difference is misleading in this context.} In Section \ref{sec3} we tie this to bubbles,\footnote{\, And to the general nontechnical arguments in \cite{AlphaOrigins}.} and also give (in a non-exhaustive fashion, solely for illustrative purposes) some empirical properties of the ratio $\kappa$ for a time horizon of about 5 quarters for a broad cross-section of U.S. equities. We find that log of market cap and sectors\footnote{\, We use the Bloomberg Industry Classification System sectors in our analysis.} appear to be relevant explanatory variables for $\kappa$, but price-to-book ratio (or its log) is not. Realized values of $\kappa$ are mostly of order 1, but have a sizable variability.

{}In Section 4 we switch gears and discuss short-horizon effects of volatility, to wit, the analog of the Heisenberg uncertainty principle in finance. A Brownian motion is obtained from a free quantum mechanical particle via the so-called Wick rotation from usual to Euclidean time. The quantum mechanical position $x(t)$ is mapped to the Brownian motion process $W_t$ and the velocity $v(t)$ is mapped to its derivative $dW_t/dt$. We discuss how to construct the analogs of quantum mechanical operators in finance and how to derive (using the binary tree) the uncertainty principle and explain the key role of {\em time ordering} therefor. We briefly conclude in Section \ref{sec5}.

\newpage
\section{A Binary Tree Model}\label{sec2}

{}In this section we will discuss a simple binary tree model in discrete time, which we will use to illustrate some long-horizon properties of stocks. Our initial time will be $t_{init} = t_0 = 0$, and our final time will be $t_{fin} = t_N = T$. Let $t_n = n\tau$, $n = 0,1,\dots,N$, be $N+1$ discrete points in time, where $\tau = T/N$. We will take $N\gg 1$, so $\tau \ll T$. Now consider a process $W_n$ where for each $0 \leq n < N$, starting from its value $W_n$, independently of the history up to time $t_n$, the process can only take two values at time $t_{n+1}$, to wit, $W_{n+1} = W_n + \sqrt{\tau}$ with probability $p$, or $W_{n+1} = W_n - \sqrt{\tau}$ with probability $q = 1 - p$. Without loss of generality, let us take $W_0 = 0$. Then the average (i.e., expected) value of $W_n$ is $\mathbb{E}(W_n) = (p - q)\sqrt{\tau}n$. For $p=q=1/2$ the process $W_n$ has vanishing expectation, i.e., it is a driftless random walk. Otherwise, if $p={1\over 2}\left[1 + \mu\sqrt{\tau}\right]$, where $\mu\neq 0$, we have $\mathbb{E}(W_n) = \mu n \tau = \mu t_n$, so $\mu$ is the drift. On the other hand, since $X_n = W_n - W_{n-1}$ is independent of $W_{n-1}$, it follows that the variance $\mbox{Var}(W_n) = \mathbb{E}(W_n^2) - [\mathbb{E}(W_n)]^2 = n \mbox{Var}(X_1) = t_n\left[1 - \mu^2\tau\right]$. In the continuous time limit, where we take $N\rightarrow \infty$ and $\tau\rightarrow 0$ while keeping $T$ fixed, $W_n$ is nothing but a Brownian motion with unit volatility and drift equal $\mu$.

{}The process $W_n$ takes both positive and negative values and is therefore unsuitable for modeling stock prices. To circumvent this, let us consider the process
\begin{equation}\label{stock.exp}
 S_n = S_0\exp(\sigma W_n)
\end{equation}
Here $\sigma$ is constant log-volatility. The expected value
\begin{equation}\label{exp.S}
 \mathbb{E}(S_n) = S_0\left[p\exp(\sigma\sqrt{\tau}) + q\exp(-\sigma\sqrt{\tau})\right]^n = S_0\exp(\nu t_n)
\end{equation}
where the drift
\begin{equation}
 \nu = {1\over\tau} \ln\left[\cosh(\sigma\sqrt{\tau}) + \mu\sqrt{\tau}\sinh(\sigma\sqrt{\tau})\right]
\end{equation}
When $\tau\ll 1/\sigma^2$ (continuous time limit), we have $\nu \approx \sigma\mu + {1\over 2}\sigma^2$.

\subsection{Example: No Drift}

{}So, suppose the drift $\nu = 0$, i.e., the expected value $\mathbb{E}(S_n) = S_0$ is independent of time $t_n$. This process is known as a martingale. Na\"ively, holding a stock whose price is described by such a driftless process might appear innocuous: on average, its value remains constant. However, this is an illusion rooted in the fact that this process is strictly positive and log-normally distributed.\footnote{\, In the continuum limit, that is.} Indeed, $\nu = 0$ invariably implies that $\mu < 0$, i.e., $W_n$ has a negative drift, or, equivalently, the probability $q$ of $W_n$ taking a step down (i.e., that $W_{n+1} = W_n -\sqrt{\tau}$) is greater than the probability $p$ of $W_n$ taking a step up (i.e., that $W_{n+1} = W_n +\sqrt{\tau}$): $q > p$. Therefore, as $T\rightarrow \infty$, invariably $W_N\rightarrow -\infty$ with probability 1 and our stock price goes to 0 with probability 1. So, holding a stock like this is a sure way to lose your money!

\subsection{Lower Bound on the Drift}

{}Based on the foregoing, it is clear that, to avoid our stock losing its value in a long run, we must have $\mu \geq 0$. This translates into a lower bound on the drift
\begin{equation}
 \nu\geq \nu_* = {1\over\tau} \ln\left[\cosh(\sigma\sqrt{\tau})\right]\approx {\sigma^2\over 2}
\end{equation}
where the last approximate equality holds for small $\tau\ll 1/\sigma^2$. So, it makes little financial sense to hold a stock long term unless its drift $\nu$ exceeds the lower bound $\nu_*$, despite the fact that on average the stock price goes up for any positive $\nu$. This is because ``on average" is not the same as ``most probably" when we have a skewed probability distribution -- which, in this case, is log-normal (in the limit $\tau\rightarrow 0$). Put differently, the expected value (\ref{exp.S}) is ``misleading" in the sense that it receives large contributions -- owing to the exponential nature of $S_n$ -- from improbable paths.

{}In this regard, we can define ``effective drift" $\nu_{eff} = \nu - \nu_*$:
\begin{equation}
 \nu_{eff} = {1\over\tau} \ln\left[1 + \mu\sqrt{\tau}\tanh(\sigma\sqrt{\tau})\right]\approx \sigma\mu
\end{equation}
The meaning of $\nu_{eff}$ is that it factors out the skewed nature of the probability distribution (log-normal). Looking at (\ref{stock.exp}), it is clear that asymptotically, when $t_n$ is large, most probable paths $S_n = {\widetilde S}_n \exp(\phi_n)$ are around the deterministic ``path" ${\widetilde S}_n = S_0\exp(\sigma\mathbb{E}(W_n)) = S_0\exp(\sigma\mu t_n)$ with the subleading fluctuations $|\phi_n| = {\cal O}(\sigma\sqrt{t_n})$ (assuming $|\mu|\gsim \sigma$, which is the case, e.g., if we have $\nu=0$).

\subsection{Including Dividends}

{}Above we assumed that our stock pays no dividends. What if it does? Na\"ively, applying the Modigliani-Miller theorem -- or, rather, its part pertinent to dividends \cite{MM} -- it might appear that dividends should not affect our discussion above. The standard argument would be that if the stock pays dividends, if we reinvest them by purchasing more stock as dividends are paid, this is equivalent to the stock not paying dividends, so dividends should make no difference. However, there is a subtlety here when we consider the long-run value of our stock.

{}Thus, if we reinvest dividends, then it is indeed the case that the situation is exactly as though the stock does not pay dividends. In this case the long-run value of the stock is zero with probability 1 unless the drift $\nu\geq \nu_*$, where $\nu$ is defined as above, irrespective of dividends. However, if we do no reinvest and instead deposit the dividends paid in a riskless savings account (which we simply assume to pay no interest or incur any fees), then even if the stock price goes to zero in the long run, we still end up with some cash. Let us quantify this based on our binary tree.

{}For the sake of simplicity, let us assume that on each date $t_n$, $n>0$, the stock pays a dividend, which is a fixed fraction $d$ of the stock price. I.e., each $t_n$ is an ex-dividend date. We can model our stock price via
\begin{equation}
 S_n = [1-d]^n S_0\exp(\sigma W_n)
\end{equation}
The cumulative dividends $D_n$ received through the date $t_n$ are given by the sum
\begin{equation}\label{divs}
 D_n = d \sum_{m=1}^n [1-d]^{m-1} S_0\exp(\sigma W_m)
\end{equation}
We can readily compute the expected value $\mathbb{E}(D_n)$. However, just as $\mathbb{E}(S_n)$, $\mathbb{E}(D_n)$ is ``misleading" in the sense that it does not take into account the skewed nature of the probability distribution, i.e., it does not correspond to the most probable paths. We can define an estimate ${\widetilde D}_n$ of $D_n$ corresponding to most probable paths by simply substituting $W_m$ via $\mathbb{E}(W_m)=\mu t_m$ in (\ref{divs}):
\begin{equation}\label{divs1}
 {\widetilde D}_n = d \sum_{m=1}^n [1-d]^{m-1} S_0\exp(\sigma\mu\tau m) = d\exp(\sigma\mu\tau)~ {{1-\exp(-\eta\tau n)}\over{1-\exp(-\eta\tau)}}~S_0
\end{equation}
where
\begin{equation}
 \eta = -{1\over\tau}\ln(1-d) - \sigma\mu
\end{equation}
Let\footnote{\, Note that $\ln(1-d)$ must be ${\cal O}(\tau)$ as $\tau\rightarrow 0$ to get the correct continuous dividend rate $\delta$.} $d = \tau\delta$. Then, in the limit where $\eta\tau\ll 1$ and $\eta t_n \gg 1$, we have (assuming $\sigma\mu < \delta$):
\begin{equation}
 {\widetilde D}_n \approx {\delta\over{\delta - \sigma\mu}}~S_0 \approx {\delta\over{\delta - \nu_{eff}}}~S_0
\end{equation}
When $\mu = 0$ (i.e., $\nu_{eff} = 0$), we have ${\widetilde D}_n = S_0\left[1-(1-d)^n\right]$, so, asymptotically, ``most probably" we (approximately) get our original investment $S_0$ back via dividend payments. When $\mu < 0$ (i.e., $\nu_{eff} < 0$), we get back a fraction of $S_0$. And when $\mu > 0$ (i.e., $\nu_{eff} > 0$), we make money just from dividends. It is instructive to contrast this with the expected value
\begin{equation}\label{divs2}
 \mathbb{E}(D_n) = d \sum_{m=1}^n [1-d]^{m-1} S_0\exp(\nu\tau m) = d\exp(\nu\tau)~ {{1-\exp(-\xi\tau n)}\over{1-\exp(-\xi\tau)}}~S_0\approx {\delta\over{\delta - \nu}}~S_0
\end{equation}
where
\begin{equation}
 \xi = -{1\over\tau}\ln(1-d) - \nu \approx \delta - \nu
\end{equation}
The expected value $\mathbb{E}(D_n)$ is ``misleading" in the sense that it is higher than the estimate ${\widetilde D}_n$ corresponding to most probable paths. The upshot is that dividends do make a difference if we do not reinvest them but stash the cash instead. This, again, is owing to the difference between ``on average" and ``most probably".\footnote{\, A complementary way to think about this is in the context of option pricing. As above, let us assume zero interest rate. Consider a European call option with the strike price $k=S_0$ and maturity $T$. Under the Black-Scholes pricing model \cite{BS}, the price of this option is $V_c = S_0\mbox{erf}(\sigma\sqrt{T}/2\sqrt{2})$. As $T\rightarrow \infty$, $V_c \rightarrow S_0$. So, despite the fact that the stock process under the risk-free probability measure (which is used in computing the option price) has a constant expectation $\mathbb{E}(S_t)\equiv S_0$ (in continuous time $t$), the option to purchase the stock at price $S_0$ in the distant future is not zero (the ``na\"ive" answer) but the full price $S_0$. This is because under this risk-free measure the stock price goes to zero with probability 1 as $T\rightarrow \infty$.}

\section{What Does This Have to Do with Bubbles?}\label{sec3}

{}A short answer is: ``Everything!" Consider a non-dividend stock. The na\"ive thinking that so long as the drift $\nu \geq 0$ we do not lose money long term because its expectation $\mathbb{E}(S_n)$ does not decrease is misguided. As we argue above, unless $\nu_{eff} \geq 0$, the most probable outcome is that the stock price will go to zero long term. I.e., investing in a non-dividend stock makes little financial sense unless $\nu_{eff} > 0$, or, equivalently, $\nu > \nu_* \approx {1\over 2}\sigma^2$. So, where can this substantial drift in the stock price come from?

{}Throughout, here we assume a free market, where prices are determined by supply and demand. There are two sources of supply and demand: investors and the company that issued the stock itself. By ``investors" we broadly mean market participants, including companies other than the issuing company, which can affect the stock price via mergers and acquisitions, tender offers, etc. The investors can increase the price of the stock by purchasing shares, or lower it by selling or shorting shares. The issuing company can lower the price of its stock by floating more shares,\footnote{\, Splits and dividends in the zeroth approximation should not affect adjusted stock prices, albeit there can be an ``irrational"/behavioral effect beyond the zeroth approximation. For the sake of simplicity we ignore it here as it does not affect our main point.} or increase it via, e.g., buybacks. If the issuing company does nothing of the kind, i.e., does not offer or purchase shares, then the price is set by the investors. The investors may or may not be rational, and may or may not base their decisions on valuations using the company's realized and projected earnings. Either way, since no cash is flowing from the company into the stock, i.e., since the company is not using its earnings to affect the stock price directly via, e.g., buybacks, the price and any drift are based purely on the investors' perception and interpretation of the information about the company's earnings and other possibly pertinent information.

{}Now, unless $\nu \geq \nu_*$, the likely long-run price of the stock is zero. So, assuming no dividends, buybacks, etc., if $\nu <\nu_*$, we have a bubble. In plain English, investors are holding cut paper. In fact, it is worse. It has all the elements of a Ponzi scheme.\footnote{\, According to the SEC (United States Securities and Exchange Commission) website http://www.sec.gov/answers/ponzi.htm, ``A Ponzi scheme is
an investment fraud that involves the payment of purported returns to existing investors from
funds contributed by new investors." SEC further clarifies: ``With little or no legitimate earnings,
Ponzi schemes require a consistent flow of money from new investors to continue. Ponzi
schemes tend to collapse when it becomes difficult to recruit new investors or when a large
number of investors ask to cash out."} And this applies even if $\nu \geq \nu_*$. Indeed, since the issuing company does not reward the investors by returning money to them in any way, shape or form, the current investors' hopes for any future gains are solely based on the future investors continuing the current investors' perception and interpretation of the earnings, etc.

{}So, unless $\nu \geq \nu_*$, we have a bubble. And, absent any cash inflow into the stock from the issuing company, it is the investors who must pony up the ``premium" $\nu_*$ in the drift $\nu$, which ``premium" is solely due to the stock's riskiness. Let
\begin{equation}\label{kappa.approx}
 \kappa = {\nu \over \nu_*} \approx {2\nu\over\sigma^2}
\end{equation}
Stocks with the expected value of $\kappa < 1$ are to be avoided (or even shorted). The higher the expected value of $\kappa$, the safer the bet to invest into the stock. The ratio $\kappa$, which apparently has not been named in the finance context,\footnote{\, Assuming vanishing interest rates, the ratio $\nu/\sigma$ is known as the market price of risk. When properly normalized (it has the dimension of $1/\sqrt{t}$), it coincides with the Sharpe/information ratio.} is dimensionless. Therefore, it is independent of the time scale based on which it is computed.

\subsection{Empirical Properties}\label{sub.empirical}

{}It is instructive to look at some empirical properties of the ratio $\kappa$ for equities, including any explanatory variables that may have statistically significant contributions into it. For longer horizons, one may expect, e.g., price-to-book ratio (P/B) to contribute to it. Here our intent is not to do an exhaustive empirical study on various time horizons. Instead, we simply discuss an illustrative example.

{}We take pricing data (closing prices fully adjusted for splits and dividends) from 2014-03-18 to 2015-06-30 (both inclusive, total of 325 trading days) for U.S. listed common stocks and class shares\footnote{\, More precisely, we keep only one ticker per company, which can be a class share.} (no OTCs, preferred shares, etc.) with no NAs in closing prices,\footnote{\, We use both raw close prices and close prices fully adjusted for splits and dividends.} market cap data, and BICS (Bloomberg Industry Classification System) sector assignments as of 2015-06-30,\footnote{\, The data was downloaded pre-open from http://finance.yahoo.com on 2015-07-01.} and no NAs in the market cap data as of 2014-03-18.\footnote{\, The market cap data as of 2014-03-18 was downloaded pre-open on 2014-03-19.} This universe contains 4416 tickers. Based on this data we compute the time series of daily close-to-close returns $R_{is}$, an $N \times K$ matrix, where $i=1,\dots,N$, $s=1,\dots,K$, $N=4416$, and $K = 324$. We use these returns to compute $\kappa_i$ (for each stock) via (\ref{kappa.approx}). Thus,
\begin{eqnarray}
 &&\mu_i = \mbox{Mean}(R_{is})\label{calc.mu}\\
 &&\sigma_i^2 = \mbox{Var}(R_{is})\label{calc.var}
\end{eqnarray}
where the mean and variance are serial.\footnote{\, The ``yesterday's close to today's close" returns $R_{is}$ are automatically ``time-ordered".} Alternatively, instead of $R_{is}$, we can use cross-sectionally demeaned returns
\begin{equation}\label{R.twiddle}
 {\widetilde R}_{is} = R_{is} - {1\over N}\sum_{j=1}^N R_{js}
\end{equation}
in (\ref{calc.mu}) and (\ref{calc.var}). I.e., here we take returns w.r.t. to an equally weighted ``market" benchmark. Yet another choice (out of myriad others) is to use returns w.r.t. a log-cap weighted ``market" benchmark:
\begin{equation}\label{R.hat}
 {\widehat R}_{is} = R_{is} - {\sum_{j = 1}^N \ln\left(C_j\right) R_{js} \over \sum_{j = 1}^N \ln\left(C_j\right)}
\end{equation}
where $C_i$ is the market cap as of 2014-03-2018, so it is out-of-sample. The results are summarized in Table \ref{table.kappa}, where for comparison purposes (and keeping in mind ``shortcomings" of cap weighted benchmarks) we also give the results for $\kappa_i$ computed based on ${\overline R}_{is}$ defined as in (\ref{R.hat}) but with $\ln(C_j)$ replaced by $C_i$, i.e., these are returns w.r.t. a cap weighted ``market" benchmark. Coincidentally, the mean value of $\kappa_i$ based on ``vanilla" returns $R_{is}$ came out close to 1 in this particular sample.

{}So, the values of $\kappa_i$ are mostly of order 1. What do they depend on? One good guess is that log of market cap should be a good explanatory variable. As mentioned above, another explanatory variable one might be tempted to try is P/B (or its log); however, it turns out to be a poor predictor of $\kappa_i$. On the other hand, sectors are much better predictors. To see this, we run cross-sectional regressions of the form:
\begin{equation}
 \ln(\kappa_i) = a + \sum_{i=1}^K b_A~U_{iA} + \varepsilon_i
\end{equation}
Here, $a$ is the regression coefficient for the intercept; $b_A$ are the regression coefficients for the $K$ explanatory variables $U_A$ (each of which is an $N$-vector), which can be, for example, $\ln(C_i)$, binary sector columns $\Omega_{i\alpha}$ (where $\Omega_{i\alpha} = 1$ if the ticker labeled by $i$ belongs to the sector labeled by $\alpha$; otherwise, $\Omega_{i\alpha} = 0$), P/B (or its log), etc. (and their number $K$ can be one or more); and $\varepsilon_i$ are the regression residuals.

{}Table \ref{table.reg.cap} summarizes the regression results when we have a single explanatory variable $\ln(C_i)$ (recall that market cap $C_i$ is taken as of 2014-03-18, so it is out-of-sample) plus the intercept.\footnote{\, We use $\kappa_i$ based on $R_{is}$. Regressions for ${\widetilde R}_{is}$, ${\widehat R}_{is}$ and ${\overline R}_{is}$ are qualitatively similar.} Table \ref{table.reg.cap.sec} summarizes the regression results when we have 11 explanatory variables, $\ln(C_i)$ plus 10 BICS sectors.\footnote{\, Our universe of 4416 tickers falls into 177 BICS sub-industries (most granular level) and 48 BICS industries (less granular level). The 10 BICS sectors correspond to the least granular level.} There is no need to include the intercept in this regression: the intercept is subsumed in the binary loadings matrix $\Omega_{i\alpha}$ for the BICS sectors as each ticker belongs to 1 and only 1 sector and $\sum_{\alpha=1}^{10} \Omega_{i\alpha} \equiv 1$. Table \ref{table.reg.cap.sec} suggests that the BICS sectors are relevant explanatory variables. Their addition sizably improves the overall R-squared (compared with including only the intercept, which is an explanatory variable corresponding to the overall ``market" -- see Table \ref{table.reg.cap}), and also the t-statistic for the market cap variable.

{}Next, we throw P/B into the mix. The book data is as of 2014-03-18,\footnote{\, This data was downloaded pre-open on 2014-03-19. As usual, book = book value per share.} and has no NAs for our universe of 4416 stocks. However, 80 tickers have 0 book value, and 159 tickers have negative book value. We exclude such tickers from our regressions below (so we can take log of P/B).\footnote{\, The ticker counts in the BICS sectors before and after this removal are in Table \ref{table.sector.counts}.} The regression results are summarized in Tables \ref{table.reg.cap.pb} and \ref{table.reg.cap.sec.pb}, which unequivocally suggest that P/B is not a good predictor of $\kappa_i$, at least on the time horizon tested (roughly, 5 calendar quarters). On the other hand, market cap and the 10 BICS sectors appear to be much better predictors of $\kappa_i$ (Table \ref{table.reg.cap.sec}).

\section{Heisenberg's Uncertainty Principle}\label{sec4}

So, the foregoing story is that, because stock prices are i) positive-definite and ii) intrinsically volatile, it is misleading to forecast their {\em long-term} behavior by looking at their expected returns as ``on average" is {\em not} the same as ``most probably", and the latter is what matters long-term. Above we discussed this on a simple example where both volatility and drift are constant. Making these time-dependent or even nondeterministic simply complicates the math, but the key principle is still valid.

{}What about the short-horizon behavior? At short horizons, due to locality considerations, the positive-definiteness of prices is unimportant. However, volatility yields a neat effect: the analog of Heisenberg's uncertainty principle for finance.

\subsection{Uncertainty Principle in Quantum Mechanics}

{}To simplify the discussion, let us consider a classical particle in one dimension. Its position is described by the coordinate $x(t)$, which depends on time $t$. Its velocity is given by $v(t) = \dot{x}(t) = dx(t)/dt$. Both $x(t)$ and $v(t)$ are real-valued. We can also define the momentum $p(t) = m~v(t)$, where $m$ is the mass of the particle. In the absence of any external forces the momentum is conserved (i.e., its time derivative vanishes, $\dot{p}(t) = 0$), so the velocity is constant. Everything is nicely deterministic.

{}However, quantum mechanics puts this idyllic picture on its head. In the so-called Heisenberg representation (a.k.a. the Heisenberg picture), the variables $x(t)$ and $p(t)$ are no longer real-valued but are replaced by operators ${\widehat x}(t)$ and ${\widehat p}(t)$ whose eigenvalues take real values. Furthermore, these operators do not commute and we have the celebrated Heisenberg uncertainty principle \cite{Heisenberg}
\begin{equation}\label{xp.comm}
 {\widehat x}(t)~{\widehat p}(t) - {\widehat p}(t)~{\widehat x}(t) = i\hbar
\end{equation}
where $\hbar$ is the (reduced) Planck's constant. This implies that both the position and the momentum (or, equivalently, the velocity) of the particle cannot be simultaneously known with infinite precision \cite{Kennard}, \cite{Weyl} (also see below):
\begin{equation}\label{unc.ineq}
 \sigma_x~\sigma_p \geq {\hbar\over 2}
\end{equation}
where $\sigma_x$ and $\sigma_p$ are standard deviations in precisions of measurements of position and momentum in an ensemble of individual measurements on similarly prepared quantum mechanical systems. Only either $x$ or $p$ can be measured infinitely precisely.

\subsubsection{Operators and Wavefunctions}

{}We can readily construct the operators ${\widehat x}$ and ${\widehat p}$. Their form depends on the basis or representation. In the coordinate representation, where ${\widehat x} = x$ (i.e., ${\widehat x}$ acts on a space of functions of $x$ by simply multiplying such a function by $x$), we have
\begin{equation}\label{mom}
 {\widehat p} = -i\hbar~{\partial\over\partial x}
\end{equation}
so we have (\ref{xp.comm}). The eigenfunction of ${\widehat p}$ (a.k.a. a wavefunction) corresponding to an eigenvalue $p$ is given by (up to a normalization constant)
\begin{equation}\label{psi.p}
 \psi_p(x) = \exp(ipx/\hbar)
\end{equation}
This wavefunction corresponds to a free quantum particle with a fixed momentum $p$. However, its position $x$ is utterly arbitrary consistently with the uncertainty principle. Conversely, let us consider a particle with a fixed position at $x_*$. This must be the sole eigenvalue of the operator ${\widehat x} = x$. The corresponding wavefunction is (up to a normalization constant)\footnote{\, Instead of the $\delta$-function, we can take its Gaussian (or some other) approximation $\exp(-(x-x_*)^2/2\epsilon^2)/\sqrt{2\pi}\epsilon$ with $\epsilon\rightarrow 0$. I.e., this is a way of regularizing the $\delta$-function.}
\begin{equation}
 \psi_{x_*}(x) = \delta(x - x_*)
\end{equation}
We can rewrite it via a Fourier transform:
\begin{equation}
 \psi_{x_*}(x) = {1\over 2\pi\hbar}\int_{-\infty}^\infty dk~\exp(ik(x - x_*)/\hbar)
\end{equation}
When the operator (\ref{mom}) acts on this wavefunction, the resultant function receives contributions from all momenta $k\in {\bf R}$:
\begin{equation}
 {\widehat p}~\psi_{x_*}(x) = {1\over 2\pi\hbar}\int_{-\infty}^\infty dk~k~\exp(ik(x - x_*)/\hbar)
\end{equation}
which, again, is consistent with the uncertainty principle.

\subsection{Uncertainty Principle in Finance}

{}A free quantum mechanical particle is not exactly what we need in finance. Thus, there are no imaginary (or complex) numbers in finance. This is just as well: Brownian motion, which appears in finance, is not equivalent to a free quantum mechanical particle in usual time $t$ but in {\em Euclidean} time $t_E$, which is obtained via the so-called Wick rotation: $t = -it_E$ \cite{Wick}.\footnote{\, For a detailed recent discussion, see, e.g., \cite{Path} and references therein.} In the following, for notational convenience, we will drop the subscript $E$ from $t_E$. Under the Wick rotation the velocity $v \rightarrow iv$, and so the momentum $p\rightarrow ip$. Therefore, in Euclidean quantum mechanics we expect the following commutation relation (instead of (\ref{xp.comm})):
\begin{equation}\label{xp.comm.e}
 {\widehat x}(t)~{\widehat p}(t) - {\widehat p}(t)~{\widehat x}(t) = \hbar
\end{equation}
and, in the coordinate representation,\footnote{\, A technical comment: the operator (\ref{mom}) is Hermitian on the space of plane-wave normalizable wavefunctions, while (\ref{mom.e}) is not (it is anti-Hermitian). However, unlike in quantum mechanics, in finance there is no ``wavefunction", nor is there a requirement that (\ref{mom.e}) act on plane-wave normalizable functions. The eigenvalues of (\ref{mom.e}) are real when acting on a proper set of eigenfunctions (which are not plane-wave normalizable). Certain subtleties arise in this regard (see below).}
\begin{equation}\label{mom.e}
 {\widehat p} = -\hbar~{\partial\over\partial x}
\end{equation}
In fact, for our purposes here, it will be more convenient to work with the velocity operator ${\widehat v} = {\widehat p} / m$, so we have
\begin{equation}\label{xv.comm}
 {\widehat x}(t)~{\widehat v}(t) - {\widehat v}(t)~{\widehat x}(t) = \zeta
\end{equation}
Where $\zeta = \hbar / m$. However, there is no $\hbar$ or mass $m$ in finance. So, in the finance context $\zeta$ must have some other interpretation. As we will see, it does indeed.

{}A more ``disconcerting" feature of the uncertainty principle (\ref{xv.comm}) may appear to be that in finance we are not accustomed to thinking about things in terms of operators. Thus, in continuous time, we have real-valued processes such as a Brownian motion $W_t$. So, why would things not commute? The answer is neat and stems from how the commutation relation (\ref{xv.comm}) is interpreted in (Euclidean) quantum mechanics (including how operators are defined in the financial context).

{}It is all about {\em time ordering}. The operators in (\ref{xv.comm}) are acting on a space of some functions. And for two operator ${\widehat A}(t)$ and ${\widehat B}(t)$, the product ${\widehat A}(t){\widehat B}(t)$ is understood as being time-ordered:
\begin{equation}
 {\widehat A}(t){\widehat B}(t) = \lim_{\eta \downarrow 0} {\widehat A}(t){\widehat B}(t - \eta)
\end{equation}
I.e., the operator acting first is {\em always} in the past w.r.t. to operator acting second. This is how the arrow of time is intrinsically engraved in quantum mechanics -- and, in fact, in finance (see below). Now we can readily compute the commutator (\ref{xv.comm}).

{}Let us go back to our discrete time binary tree model in Section \ref{sec2}. So, time $t = t_n = n\tau$ ($n=0,1,\dots,N$) is now discrete. We identify the coordinate $x(t)$ with our process $W_n$, i.e., $x(t_n) = W_n$. We also need to define the velocity $v(t)$. It is convenient (albeit not critical) to define it at the midpoint:
\begin{equation}
 v\left(t_n + {\tau\over 2}\right) = {{W_{n+1} - W_n}\over\tau}
\end{equation}
So, according to our time-ordering prescription, we need to evaluate
\begin{equation}\label{wv}
 G_{n,n+1} = W_{n+1}~v\left(t_n + {\tau\over 2}\right) - v\left(t_n + {\tau\over 2}\right) W_n = \tau~v^2\left(t_n + {\tau\over 2}\right)
\end{equation}
More precisely, we need to evaluate the {\em expectation} of this expression: recall that $W_{n+1} - W_n = \sqrt{\tau}$ with probability $p$ and $W_{n+1} - W_n = -\sqrt{\tau}$ with probability $q = 1 - p$ (see Section \ref{sec2}). Therefore, the expectation of (\ref{wv}) is simply\footnote{\, Note that this holds irrespective of the drift $\mu$. Recall that $p = {1\over 2}\left[1 + \mu\sqrt{\tau}\right]$.}
\begin{equation}
 \mathbb{E}(G_{n,n+1}) = 1
\end{equation}
I.e., $\zeta = 1$ in the commutator (\ref{xv.comm}) once we identify the coordinate $x(t)$ with $W_n$. Note, however, that this result persists in the continuous time limit ($\tau\rightarrow 0$, $N\rightarrow \infty$, $T=N\tau = \mbox{fixed}$), so we have the following commutation relation
\begin{equation}\label{xv.comm.fin}
 {\widehat x}(t)~{\widehat v}(t) - {\widehat v}(t)~{\widehat x}(t) = 1
\end{equation}
The reason why this is possible is that the velocity has infinite variance in the continuum limit. This is just as well. In continuous time we identify $x(t)$ with the Brownian motion process $W_t$, and $v(t)$ is identified with white noise $dW_t/dt$. The latter has infinite variance, hence the uncertainty principle (\ref{xv.comm.fin}) -- in finance.

\subsubsection{Coordinate Representation}

{}We can readily construct the velocity operator ${\widehat v}$ in the coordinate representation ${\widehat x} = x$:
\begin{equation}\label{v.op}
 {\widehat v} = -{\partial\over\partial x}
\end{equation}
Its eigenfunctions (which are {\em not} wavefunctions -- see below) are of the form
\begin{equation}\label{eigen.v}
 \psi_v(x) = \exp(-vx)
\end{equation}
where $v$ are the eigenvalues. Note that (\ref{eigen.v}) are not plane-wave normalizable. When the velocity $v$ is fixed, the coordinate $x$ is completely undetermined, consistently with the uncertainty principle. On the other hand, if we localize our Brownian motion near $x = x_*$, which we can do by considering, e.g., the Laplace distribution
\begin{equation}\label{laplace.den}
 \psi_{x_*}(x) = {1\over 2\epsilon}~\exp\left(-{|x - x_*|\over \epsilon}\right)
\end{equation}
then we have
\begin{equation}\label{v.laplace}
 {\widehat v}~\psi_{x_*}(x) = {\mbox{sign}(x - x_*)\over 2\epsilon^2}~\exp\left(-{|x - x_*|\over \epsilon}\right)
\end{equation}
and the velocity is completely undetermined in the small $\epsilon$ limit, again, consistently with the uncertainty principle. Intuitively, we can think about this as follows. If we put a Brownian particle in a box and make the box smaller and smaller, both the direction and the magnitude of the velocity of the Brownian particle become completely uncertain,\footnote{\, I would like to thank Olindo Corradini for pointing out this visualization.} same as for a (Euclidean time) quantum mechanical particle.

{}Finally, let us mention that the above derivation of the uncertainty principle using a binary tree can be ported essentially unchanged into the path integral language (see below). There, we also compute the path integral expectation of the commutator (the l.h.s. of (\ref{xv.comm.fin})), which is time-ordered inside the path integral. Upon discretization of the path integral, we then follow the same procedure as above.\footnote{\, In the path integral language (see, e.g., \cite{Baaquie}, \cite{Path}) we must specify the integration measure. This translates into our convenient choice of the midpoint in the definition of the discretized velocity above. With some care the final answer is independent of these nuances.}

\subsubsection{Inequality Form of Uncertainty Principle}

{}Above we discuss the Heisenberg uncertainty principle in the financial context in the form of the commutation relation (\ref{xv.comm.fin}). However, most people -- at least non-physicists -- associate the uncertainty principle with the inequality of the type (\ref{unc.ineq}), which follows from (\ref{xp.comm}) in quantum mechanics. Can we derive a similar inequality in the financial context starting from (\ref{xv.comm.fin})? The answer is yes, but it is tricky.

{}It would be erroneous to assume that -- in na\"{i}ve analogy with (\ref{unc.ineq}) -- the answer is simply $\sigma_x\sigma_v\geq 1/2$. This is for two reasons. First, in quantum mechanics (\ref{unc.ineq}) is derived using the notion of the wavefunction.\footnote{\, That is, in the coordinate representation, where generally a complex-valued wavefunction $\psi(x)$ describes the probability amplitude, whereas its absolute value squared $|\psi(x)|^2$ (which is real-valued) is the probability density and $\int_{-\infty}^\infty dx~|\psi(x)|^2 = 1$ (that is, for spatially localized configurations -- for non-localized configurations, such as pure momentum states (\ref{psi.p}), things are a bit trickier; however, for our purposes here we will not need to delve into such details). The expected value of a Hermitian operator ${\widehat A}$ is then computed via $\langle {\widehat A}\rangle = \int_{-\infty}^\infty dx~\psi^*(x){\widehat A}\psi(x)$, where ${\widehat A}$ is assumed to act to the right (onto $\psi(x)$), and $\psi^*(x)$ is the complex conjugate of $\psi(x)$. More generally, i.e., not in the coordinate representation, a quantum state is described by a ket vector $|\psi\rangle$ (and its conjugate bra vector $\langle\psi|$, with $\langle\psi|\psi\rangle = 1$), and operator averages are computed via $\langle\psi|{\widehat A}|\psi\rangle$ (with ${\widehat A}$ acting to the right). For our purposes here, the coordinate representation suffices.} As mentioned above, there are no wavefunctions in the financial context. Second, in quantum mechanics the operators ${\widehat x}$ and ${\widehat p}$ are Hermitian. In the financial context, they are not (at least, ${\widehat v}$ is not in the coordinate representation -- see above). In fact, $\sigma_x$ and $\sigma_v$ need to be {\em defined}.

{}Happily, we do not need wavefunctions to define $\sigma_x$ and $\sigma_v$. Let us stick to the coordinate representation. Let us assume that we have a normalizable probability density $P(x)$ for the values of $x$, i.e.,
\begin{equation}\label{norm}
 \int_{-\infty}^\infty dx~P(x) = 1
\end{equation}
That is, we have a localized configuration, and
\begin{eqnarray}
 &&x_* = \int_{-\infty}^\infty dx~x~P(x)\\
 &&\sigma_x^2 = \int_{-\infty}^\infty dx~\left(x - x_*\right)^2 P(x)
\end{eqnarray}
with finite $\sigma_x^2$. Without loss of generality, we can set $x_* = 0 $ by shifting the origin.

{}What about $\sigma_v$? Note that $\psi_{x_*}(x)$ in (\ref{laplace.den}) is one particular example of the density $P(x)$. As in (\ref{v.laplace}), we can define $v(x)$ via
\begin{equation}
 {\widehat v}~P(x) = v(x)~P(x)
\end{equation}
Therefore, recalling (\ref{v.op}), we have
\begin{equation}
 v(x) = -Q^\prime(x)
\end{equation}
where $Q(x) = \ln(P(x))$ and $Q^\prime(x) = \partial Q(x)/\partial x$. Now we can define
\begin{eqnarray}\label{v_*}
 &&v_* = \int_{-\infty}^\infty dx~v(x)~P(x)\\
 &&\sigma_v^2 = \int_{-\infty}^\infty dx \left(v(x) - v_*\right)^2 P(x)
\end{eqnarray}
Note that the integrand in (\ref{v_*}) is a total derivative, so $v_* = 0$ (as $P(x\rightarrow\pm\infty)\rightarrow 0$). Therefore, we have (we set $x_* = 0$)
\begin{eqnarray}
 &&\sigma_x^2 = \int_{-\infty}^\infty dx~x^2~P(x)\\
 &&\sigma_v^2 = \int_{-\infty}^\infty dx \left[Q^\prime(x)\right]^2 P(x)
\end{eqnarray}
The Cauchy-Schwartz inequality then immediately implies that
\begin{equation}\label{CS}
 \sigma_x^2~\sigma_v^2 \geq \left[\int_{-\infty}^\infty dx~x~Q^\prime(x)~P(x)\right]^2 = 1
\end{equation}
where the last equality follows from (\ref{norm}) and that $Q^\prime(x)~P(x) = P^\prime(x)$. So, we have
\begin{equation}\label{fin.ineq}
 \sigma_x~\sigma_v \geq 1
\end{equation}
which is the (inequality version of) the uncertainty principle in the financial context.

{}Going back to our example (\ref{laplace.den}), we have $\sigma_x^2 = 2\epsilon^2$. On the other hand, $v(x) = \mbox{sign}(x)/\epsilon$, so $v(x) = 1/\epsilon$ for $x > x_*$, $v(x) = -1/\epsilon$ for $x < x_*$, $v_* = 0$, $v^2(x) = 1/\epsilon^2$, and $\sigma^2_v = 1/\epsilon^2$. Therefore, for the Laplace density (\ref{laplace.den}), we have $\sigma_x~\sigma_v = \sqrt{2}$. On the other hand, if we take the Gaussian density $P(x) = \exp(-(x-x_*)^2/2\epsilon^2)/\sqrt{2\pi}\epsilon$, we get $\sigma_x = \epsilon$, $v(x) = (x-x_*)/\epsilon^2$, so $\sigma_v = 1/\epsilon$, and the inequality (\ref{fin.ineq}) is saturated: $\sigma_x~\sigma_v = 1$. The necessary and sufficient condition for saturating the inequality in (\ref{CS}) is $xP(x) = \alpha Q^\prime(x) P(x) = \alpha P^\prime(x)$, hence the Gaussian distribution ($\alpha=\mbox{const.}$).

\subsubsection{Where Did $1/2$ Go?}

{}But wait! What happened to the factor of $1/2$ on the r.h.s. in (\ref{unc.ineq}) we had in the context of quantum mechanics?\footnote{\, Recall that when comparing quantum mechanical and financial contexts, we set $\hbar/m=1$.} It is absent in (\ref{fin.ineq}). How come? No wavefunctions!

{}In quantum mechanics we work with the wavefunctions $\psi(x)$, not probability densities $P(x)$. Thus, we have (${\overline x} = \langle x\rangle$, ${\overline p} = \langle {\widehat p}\rangle$):
\begin{eqnarray}
 &&{\overline x} = \int_{-\infty}^\infty dx~\psi^*(x)~x~\psi(x)\\
 &&\sigma_x^2 = \int_{-\infty}^\infty dx~\psi^*(x) \left(\Delta x\right)^2 \psi(x)\\
 &&{\overline p} = \int_{-\infty}^\infty dx~\psi^*(x)~{\widehat p}~\psi(x) = -i\hbar\int_{-\infty}^\infty dx~\psi^*(x)~\psi^\prime(x)\\
 &&\sigma_p^2 = \int_{-\infty}^\infty dx~\psi^*(x) \left(\Delta {\widehat p}\right)^2 \psi(x) = \int_{-\infty}^\infty dx \left|\Delta{\widehat p}~\psi(x)\right|^2\label{sig_p}
\end{eqnarray}
where $\Delta x = x - {\overline x}$, and $\Delta {\widehat p} = {\widehat p} - {\overline p}$ (and we have integrated by parts in (\ref{sig_p})).

{}The first thing to note is that generally ${\overline p}\neq 0$. Indeed, let $\psi(x) = \rho(x)\exp(i\phi(x))$, where $\rho(x)$ and $\phi(x)$ are real. Then
\begin{eqnarray}\label{rho.norm}
 &&\int_{-\infty}^\infty dx ~\rho^2(x) = 1\\
 &&{\overline p} = \hbar \int_{-\infty}^\infty dx ~\rho^2(x)~\phi^\prime(x)
\end{eqnarray}
So, the gradient in the phase $\phi(x)$ generally yields nonzero average momentum.

{}Second, we still can apply the Cauchy-Schwartz inequality:
\begin{equation}
 \sigma_x^2~\sigma_p^2 \geq \left|\int_{-\infty}^\infty dx~\psi^*(x)~\Delta x~\Delta {\widehat p}~\psi(x)\right|^2
\end{equation}
A straightforward computation yields
\begin{equation}\label{aux}
\int_{-\infty}^\infty dx~\psi^*(x)~\Delta x~\Delta {\widehat p}~\psi(x) = {i\hbar\over 2} + \hbar\int_{-\infty}^\infty dx\left(x - {\overline x}\right) \rho^2(x)~\phi^\prime(x)
\end{equation}
where the factor of $1/2$ stems from integrating by parts and using (\ref{rho.norm}). The second term on the r.h.s. of (\ref{aux}) is real, so we get the celebrated inequality
\begin{equation}\label{xp}
 \sigma_x~\sigma_p \geq {\hbar\over 2}
\end{equation}
The extra factor of $1/2$ on the r.h.s. of (\ref{xp}) arises because the normalization condition in quantum mechanics is (\ref{rho.norm}) as opposed to (\ref{norm}). Put differently, in quantum mechanics we deal with absolute values squared (of complex wavefunctions) in lieu of probability densities, whereas in finance we deal directly with probability densities. Integration by parts then produces an extra factor of $1/2$ in quantum mechanics.

\section{Concluding Remarks}\label{sec5}

{}In this note we discuss -- in what is intended to be a pedagogical fashion -- long-horizon effects of the combination of stock prices being i) positive-definite and ii) intrinsically volatile. In particular, it is misleading to forecast {\em long-term} behavior of stock prices by looking at their expected returns as ``on average" is {\em not} the same as ``most probably", and the latter is what matters long-term. This all boils down to the fact that stock prices are not normally distributed but have skewed, long-tailed (at the higher end) distributions resulting in overly weighted contributions of improbable prices into the expected price thereby creating an illusory effect. In this regard, the highly skewed world of Manhattan's real estate provides an apt analogy: looking at mean real estate prices is essentially useless as they are distorted by a relatively small number of -- ``black swans", in the language of \cite{Taleb} -- high-end properties. A nice discussion of financial bubbles can be found in \cite{Sornette}.

{}We also discuss local (short-horizon) effects of volatility (irrespective of positive-definiteness of prices), to wit, the analog of the uncertainty principle in finance. There have been prior works alluding to the uncertainty principle in the financial context;\footnote{\, See, e.g., \cite{Biane}, \cite{Powers}, \cite{SS}.} however, these were mostly of collateral nature not directly relating to the simple map between a quantum mechanical particle and a Brownian motion, with the exception of an appendix in the book \cite{Baaquie}, where the uncertainty principle is derived in a more complicated setting using the path integral formulation. Here we use just a simple binary tree, albeit, as mentioned above, in the path integral formulation the actual computation at the final step, after going through the entire path integral machinery, is essentially the same as what we discuss herein. The advantages of our simple, binary tree based derivation are: i) no need to complicate things with path integral; and ii) the connection to the Brownian motion is explicit. Furthermore, as we explain above in detail, the financial inequality (\ref{fin.ineq}), unlike its quantum mechanical counterpart (\ref{xp}), has no factor of $1/2$ (cf. \cite{Baaquie}).

\section*{Acknowledgments}

{}I would like to thank Peter Carr and Alberto Iglesias for stimulating discussions that prompted this write-up. I would like to thank Olindo Corradini for a correspondence on interpretation of the Uncertainty Principle in Euclidean Quantum Mechanics.

\begin{table}[ht]
\noindent
\caption{Summaries of the values of $\kappa_i$ for 4416 stocks based on the vanilla returns $R_{is}$, the cross-sectionally demeaned returns ${\widetilde R}_{is}$, the returns ${\widehat R}_{is}$ w.r.t. the log-cap weighted ``market" benchmark, and the returns ${\widetilde R}_{is}$ w.r.t. the cap weighted ``market" benchmark (see Subsection \ref{sub.empirical} for details). Also see Figures 1, 2, 3 and 4. 1st Qu. = 1st Quartile, 3rd Qu. = 3rd Quartile, StDev = standard deviation, MAD = mean absolute deviation.}
\begin{tabular}{l l l l l l l l l} 
\\
\hline\hline 
Quantity & Min & 1st Qu. & Median & Mean & 3rd Qu. & Max & StDev & MAD \\[0.5ex] 
\hline 
$R_{is}$ & -22.65 & -2.577 & 0.134 & 1.096 & 4.057 & 46.45 & 5.672 & 4.700\\
${\widetilde R}_{is}$ & -18.91 & -1.514 & 2.278 & 4.373 & 8.557 & 47.57 & 8.373 & 6.798\\
${\widehat R}_{is}$ & -19.13 & -1.637 & 2.059 & 4.107 & 8.184 & 47.19 & 8.232 & 6.610\\
${\overline R}_{is}$ & -31.55 & -3.491 & -0.656 & 0.270 & 3.406 & 39.58 & 6.527 & 4.885\\ [1ex] 
\hline 
\end{tabular}
\label{table.kappa} 
\end{table}

\begin{table}[ht]
\noindent
\caption{Summary for the cross-sectional regression of $\kappa_i$ over $\ln(C_i)$ with the intercept, where $C_i$ is the market cap (out-of-sample). See Subsection \ref{sub.empirical} for details.}
\begin{tabular}{l l l l l} 
\\
\hline\hline 
 & Estimate & Standard error & t-statistic & Overall \\[0.5ex] 
\hline 
Intercept & -10.598 & 0.8465 & -12.52 & \\
$\ln(C_i)$ & 0.5655 & 0.0407 & 13.88 & \\	
Mult./Adj. R-squared & & & & 0.0418~/~0.0416\\
F-statistic	& & & & 192.7\\ [1ex] 
\hline 
\end{tabular}
\label{table.reg.cap} 
\end{table}

\begin{table}[ht]
\noindent
\caption{Summary for the cross-sectional regression of $\kappa_i$ over $\ln(C_i)$ plus the 10 BICS sectors (without the intercept -- the intercept is subsumed in the binary loadings matrix $\Omega_{i\alpha}$ for the BICS sectors as $\sum_{\alpha=1}^{10} \Omega_{i\alpha} \equiv 1$), where $C_i$ is the market cap (out-of-sample). See Subsection \ref{sub.empirical} for details.}
\begin{tabular}{l l l l l} 
\\
\hline\hline 
 & Estimate & Standard error & t-statistic & Overall \\[0.5ex] 
\hline 
$\ln(C_i)$ & 0.6348 & 0.0393 & 16.13 & \\
Consumer Discr. & -11.515 & 0.8449 & -13.63 & \\
Energy & -15.881 & 0.8775 & -18.10 & \\
Health Care & -11.286 & 0.8253 & -13.68 & \\
Financials & -9.8602 & 0.8231 & -11.98 & \\
Industrials & -13.176 & 0.8422 & -15.65 & \\
Materials & -14.215 & 0.8766 & -16.21 & \\
Communications & -13.214 & 0.8998 & -14.69 & \\
Consumer Staples & -10.415 & 0.9271 & -11.23 & \\
Utilities & -12.023 & 1.0054 & -11.96 & \\
Technology & -12.546 & 0.8339 & -15.04 & \\	
Mult./Adj. R-squared & & & & 0.1692~/~0.1671\\
F-statistic	& & & & 81.57 \\ [1ex] 
\hline 
\end{tabular}
\label{table.reg.cap.sec} 
\end{table}

\begin{table}[ht]
\noindent
\caption{Ticker counts for the 10 BICS sectors. In the second column, the total number of tickers is 4416, same as in Table \ref{table.kappa}. In the third column, the total number of tickers is 4177, after removing 80 tickers with 0 book value and 159 tickers with negative book value. See Subsection \ref{sub.empirical} for details.}
\begin{tabular}{l l l} 
\\
\hline\hline 
Sector & Count & Count\\[0.5ex] 
\hline 
Consumer Discretionary & 593 & 563\\
Energy                 & 376 & 357\\
Health Care            & 572 & 511\\
Financials             & 1010 & 955\\
Industrials            & 442 & 426\\
Materials              & 297 & 284\\
Communications         & 254 & 230\\
Consumer Staples       & 177 & 171\\
Utilities              & 102 & 100\\
Technology             & 593 & 580\\ [1ex] 
\hline 
\end{tabular}
\label{table.sector.counts} 
\end{table}

\begin{table}[ht]
\noindent
\caption{Summary for the cross-sectional regression of $\kappa_i$ over $\ln(C_i)$ and $X$ with the intercept, where $C_i$ is the market cap (out-of-sample), and $X$ is P/B (price-to-book, out-of-sample) in the first sub-table, B/P (book-to-price, inverse of P/B) in the second sub-table, and log of P/B in the third sub-table. The number of tickers is 4177. See Subsection \ref{sub.empirical} for details.}
\begin{tabular}{l l l l l} 
\\
\hline\hline 
{} & Estimate & Standard error & t-statistic & Overall \\[0.5ex] 
\hline 
Intercept & -10.09 & 0.8647 & -11.67 & \\
$\ln(C_i)$ & 0.542 & 0.0416 & 13.04 & \\
P/B & $-6\cdot 10^{-5}$ & $3\cdot 10^{-4}$ & -0.224 & \\	
Mult./Adj. R-squared & & & & 0.0392~/~0.0387\\
F-statistic	& & & & 85.05\\ [1ex]

\hline 
Intercept & -10.307 & 0.9219 & -11.18 & \\
$\ln(C_i)$ & 0.5496 & 0.0431 & 12.76 & \\
B/P & 0.1007 & 0.1454 & 0.692 & \\	
Mult./Adj. R-squared & & & & 0.0393~/~0.0388\\
F-statistic	& & & & 85.27\\ [1ex]

\hline
Intercept & -10.463 & 0.8712 & -12.01 & \\
$\ln(C_i)$ & 0.5730 & 0.0426 & 13.46 & \\
$\ln(\mbox{P/B})$ &  -0.2967 & 0.0905 & -3.277 & \\	
Mult./Adj. R-squared & & & & 0.0416~/~0.0412\\
F-statistic	& & & & 90.61\\ [1ex] 
\hline
\end{tabular}
\label{table.reg.cap.pb} 
\end{table}

\begin{table}[ht]
\noindent
\caption{Summary for the cross-sectional regression of $\kappa_i$ over $\ln(C_i)$, the 10 BICS sectors and log of P/B (without the intercept -- see Table \ref{table.reg.cap.sec}), where $C_i$ is the market cap (out-of-sample), and P/B is the price-to-book ratio (also out-of-sample). See Subsection \ref{sub.empirical} for details.}
\begin{tabular}{l l l l l} 
\\
\hline\hline 
 & Estimate & Standard error & t-statistic & Overall \\[0.5ex] 
\hline 
$\ln(C_i)$ & 0.6318 & 0.0413 & 15.30 & \\
Consumer Discr. & -11.253 & 0.8656 & -13.00 & \\
Energy & -15.725 & 0.9050 & -17.38 & \\
Health Care & -10.875 & 0.8450 & -12.87 & \\
Financials & -9.5871 & 0.8565 & -11.19 & \\
Industrials & -13.065 & 0.8657 & -15.09 & \\
Materials & -14.026 & 0.9037 & -15.52 & \\
Communications & -13.047 & 0.9250 & -14.11 & \\
Consumer Staples & -10.193 & 0.9469 & -10.76 & \\
Utilities & -11.887 & 1.0341 & -11.50 & \\
Technology & -12.304 & 0.8530 & -14.42 & \\	
$\ln(\mbox{P/B})$ & -0.1660 & 0.0934 & -1.777\\
Mult./Adj. R-squared & & & & 0.1772~/~0.1748\\
F-statistic	& & & & 74.74 \\ [1ex] 
\hline 
\end{tabular}
\label{table.reg.cap.sec.pb} 
\end{table}

\clearpage
\newpage
\begin{figure}[ht]
\centerline{\epsfxsize 4.truein \epsfysize 4.truein\epsfbox{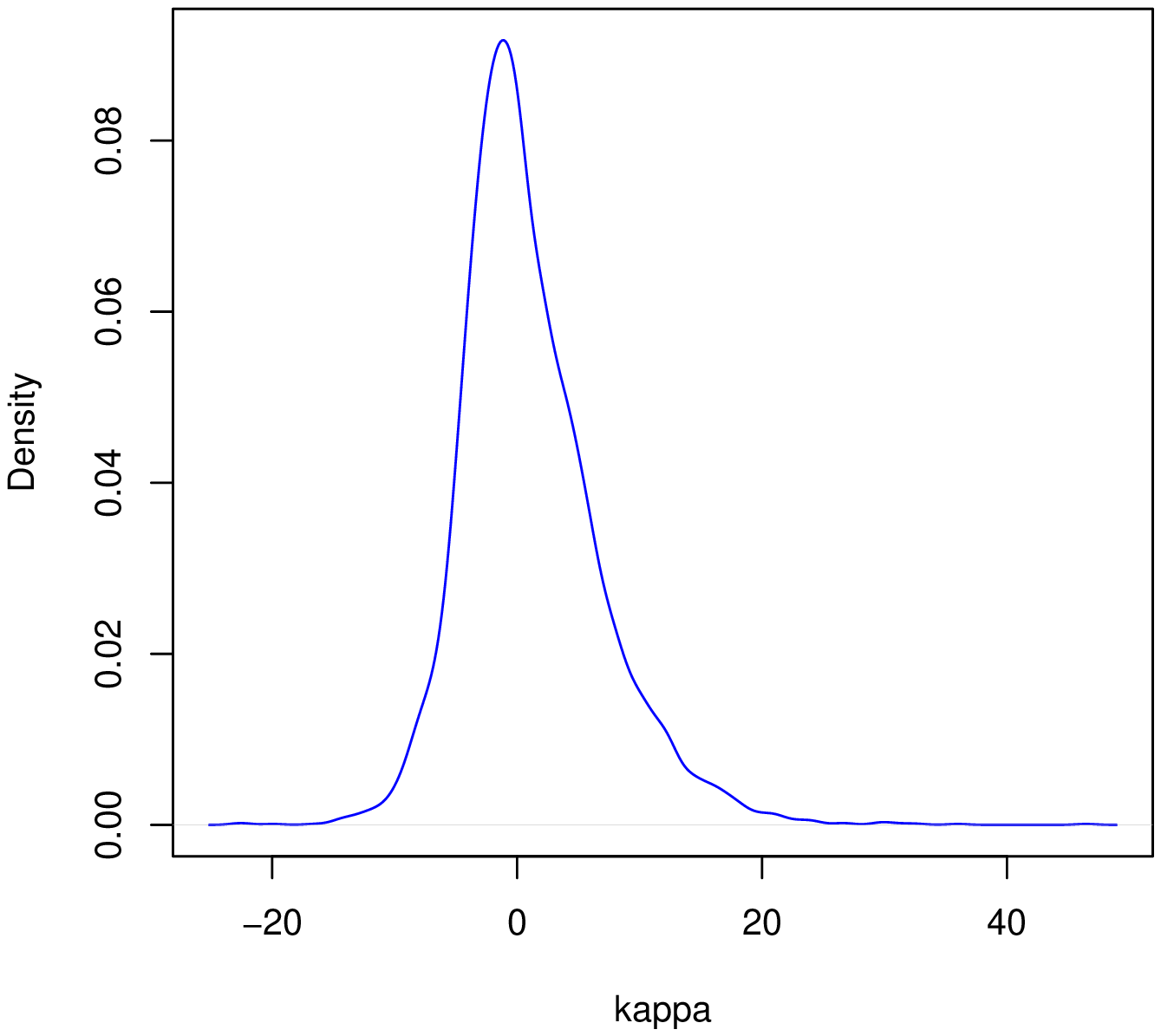}}
\noindent{\small {Figure 1. Density of $\kappa_i$ based on $R_{is}$ (see Table \ref{table.kappa}).}}
\end{figure}

\clearpage
\newpage
\begin{figure}[ht]
\centerline{\epsfxsize 4.truein \epsfysize 4.truein\epsfbox{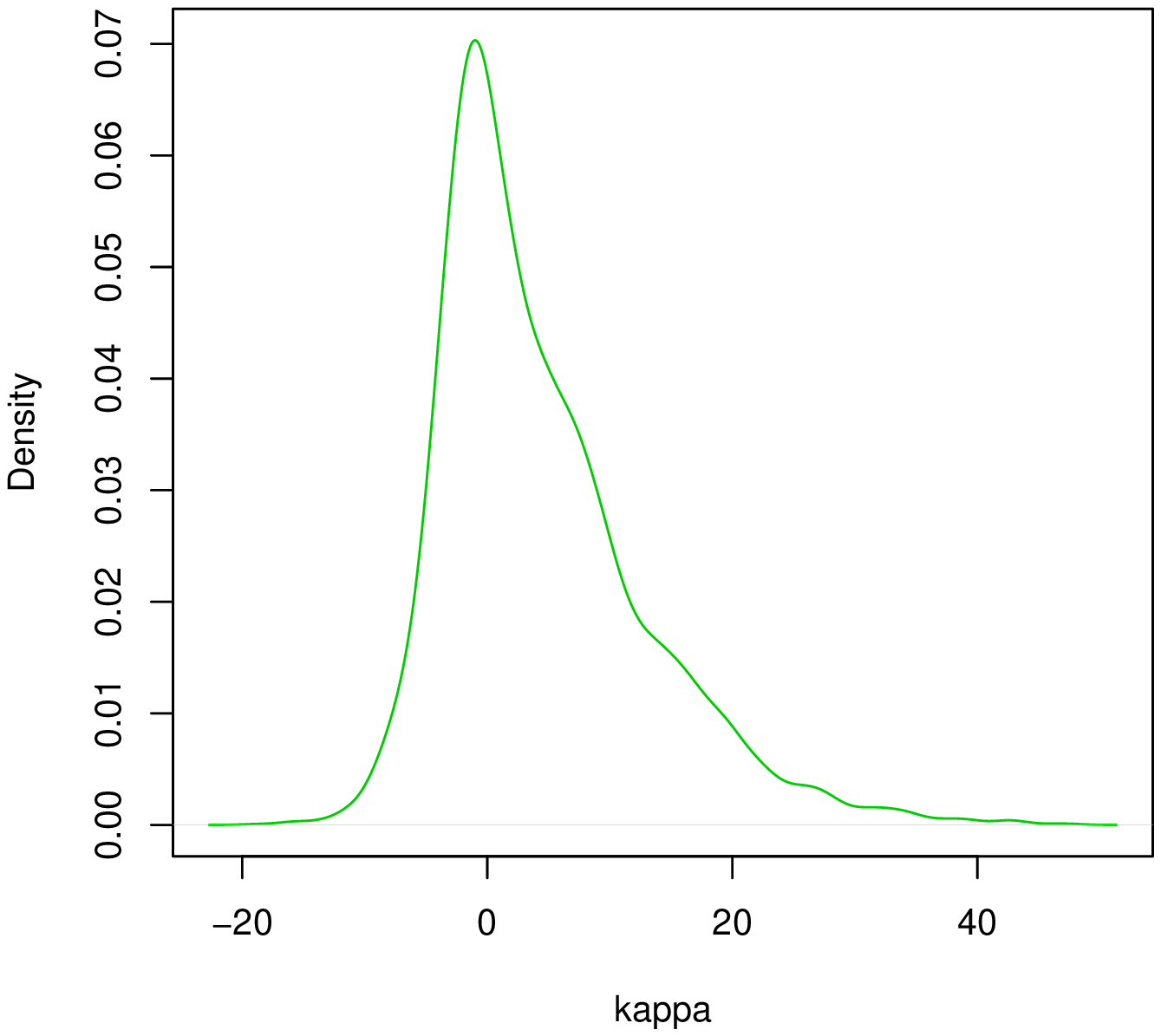}}
\noindent{\small {Figure 2. Density of $\kappa_i$ based on ${\widetilde R}_{is}$ (see Table \ref{table.kappa}).}}
\end{figure}

\clearpage
\newpage
\begin{figure}[ht]
\centerline{\epsfxsize 4.truein \epsfysize 4.truein\epsfbox{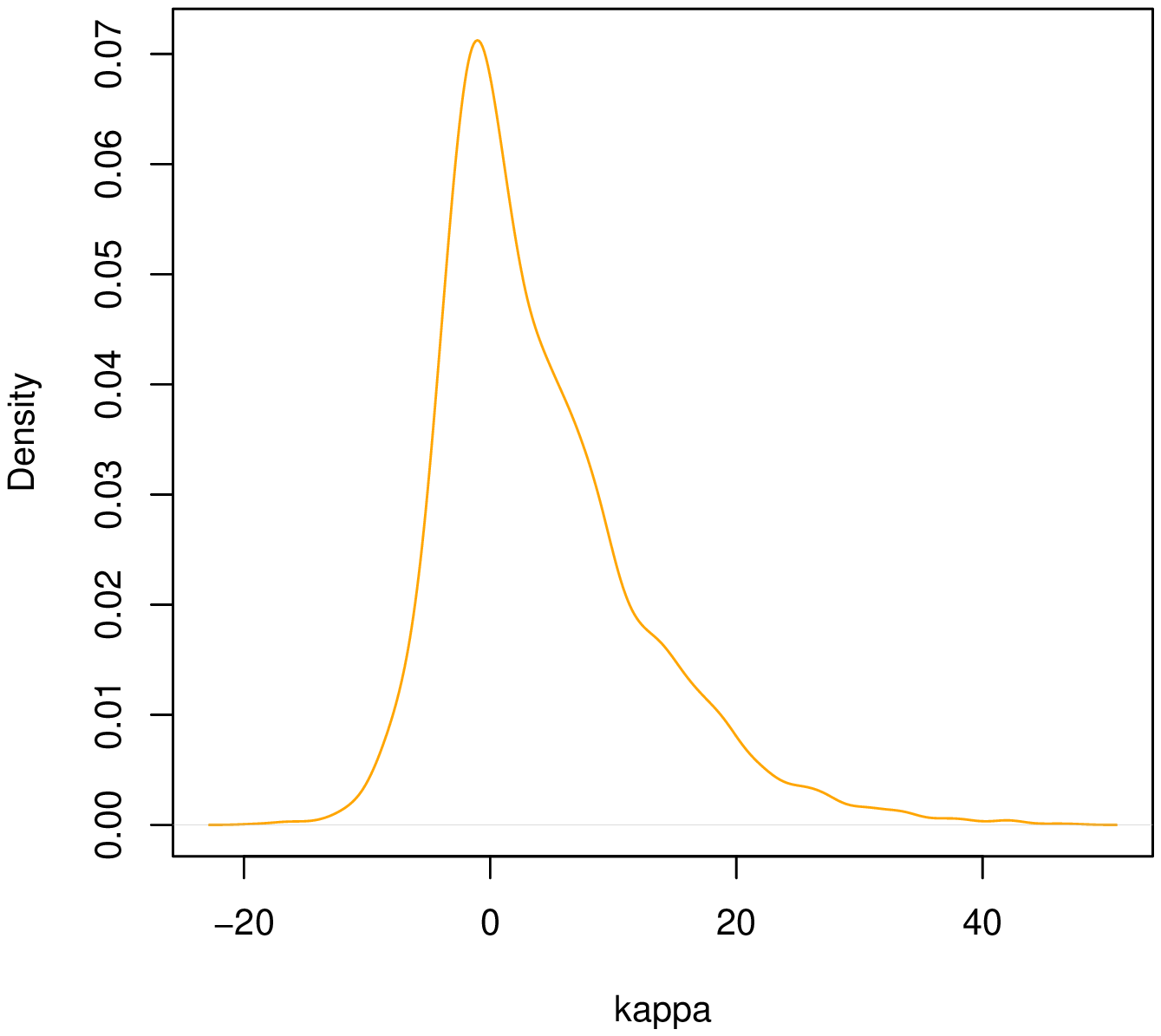}}
\noindent{\small {Figure 3. Density of $\kappa_i$ based on ${\widehat R}_{is}$ (see Table \ref{table.kappa}).}}
\end{figure}

\clearpage
\newpage
\begin{figure}[ht]
\centerline{\epsfxsize 4.truein \epsfysize 4.truein\epsfbox{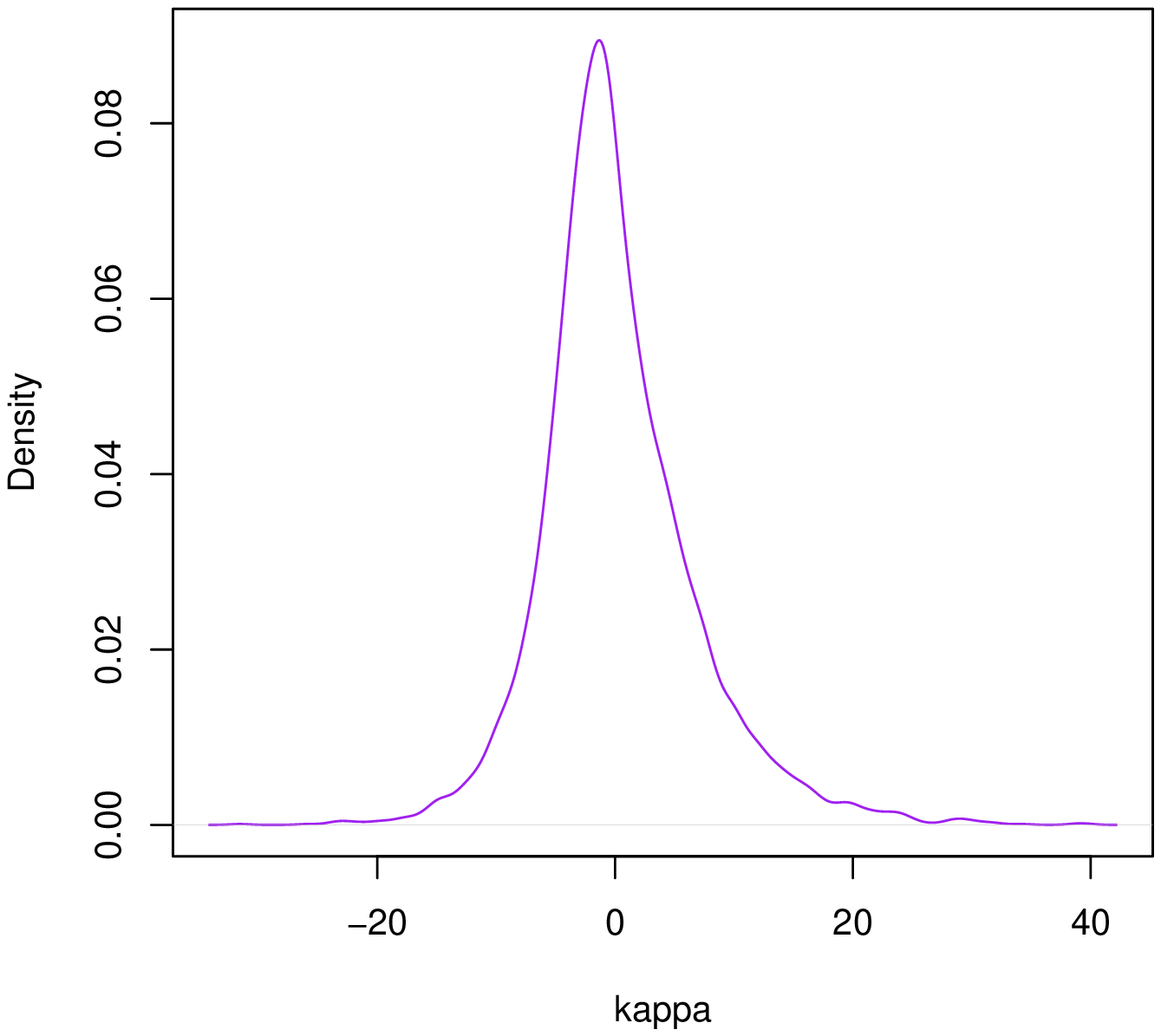}}
\noindent{\small {Figure 4. Density of $\kappa_i$ based on ${\overline R}_{is}$ (see Table \ref{table.kappa}).}}
\end{figure}

\end{document}